\documentstyle[12pt,amssym]{article}
\def\C{\mbox{$\Bbb C$}}
\def\R{\mbox{$\Bbb R$}}
\font\twentyof=msbm10 at 20pt

\font\sixteenof=msbm10 at 16pt

\def\case#1#2{{\textstyle{#1\over #2}}}
\def\sech{\mathop{\rm sech}\nolimits}
\def\cosech{\mathop{\rm cosech}\nolimits}

\title{
\hfill{\normalsize ULB/229/CQ/00/4}\\
\vspace{1cm}
sl(2, \mbox{\twentyof C}) as a complex Lie algebra and the associated
non-Hermitian Hamiltonians with real eigenvalues}

\author{B.\ Bagchi $^{a,}$\thanks{E-mail: bbagchi@cucc.ernet.in}\ , C.\ Quesne
$^{b,}$\thanks{Directeur de recherches FNRS; E-mail: cquesne@ulb.ac.be} \\
{\small \sl $^a$ Department of Applied Mathematics, University of Calcutta,} \\
{\small \sl 92 Acharya Prafulla Chandra Road, Calcutta 700 009, India}\\
{\small \sl $^b$ Physique Nucl\'eaire Th\'eorique et Physique
Math\'ematique,  Universit\'e Libre de Bruxelles,} \\ {\small \sl Campus de la
Plaine CP229, Boulevard~du Triomphe, B-1050 Brussels, Belgium}}
\date{ }
\begin{document}
\baselineskip=22pt plus 1pt minus 1pt
\maketitle

\begin{abstract}
The powerful group theoretical formalism of potential algebras is extended to
non-Hermitian Hamiltonians with real eigenvalues by complexifying so(2,1),
thereby
getting the complex algebra sl(2,\C) or $A_1$. This leads to new types of both
PT-symmetric and non-PT-symmetric Hamiltonians.
\end{abstract}

\vspace{0.5cm}



\newpage
%
%
\section{Introduction}

In recent times, the subject of quantum mechanics has been the focus of very
active research. While on the one hand, new methods have been developed which
provide deeper insights into the underpinnings of the
theory~\cite{ballentine}, on
the other, group theoretical methods have increasingly gained acceptance as an
important means towards understanding the extremely rich structure of the
underlying symmetries~\cite{alhassid, bohm}.\par
%
%
Of late, a sharp increase of interest has been noticed in searching for
non-Hermitian Hamiltonians~\cite{bender98a, bender98b, cannata, andrianov,
znojil99a, bagchi00a, znojil00, bagchi00b, znojil99b}. Although the history of
complex potentials is old~\cite{feshbach} especially in relation to scattering
problems, Bender and Boettcher~\cite{bender98a}, a few years ago, revived
interest
in complex potentials by restricting a non-Hermitian Hamiltonian to be
PT-symmetric. In this way, they showed that it is possible to derive new
infinite
classes of PT-invariant systems whose spectrum is real. Subsequently the idea of
PT symmetry has been pursued by several authors~\cite{bender98b,
cannata, andrianov, znojil99a, bagchi00a, znojil00, bagchi00b, znojil99b},
who have
obtained different kinds of potentials with real eigenvalues. These include the
quasi-solvable type~\cite{bender98b, bagchi00b} and supersymmetry-inspired ones
too~\cite{andrianov, bagchi00a, znojil00}.\par
%
%
In this letter we propose an sl(2,\C) potential algebra as a complex Lie
algebra for
the Schr\"odinger equation to study non-Hermitian systems from a group
theoretical point of view. Adopting a most general differential realization
of the
sl(2,\C) algebra, we demonstrate how new complex potentials can be generated
which are not necessarily PT symmetric but possess common real eigenvalues.
However, a subclass of our potentials does turn out to respect PT symmetry. As
with the case of so(2,1)~\cite{wu, englefield}, here also we find possible to
classify our results into various types of solutions. Indeed the main
spirit of the
realization of the potential algebra so(2,1) persists in our scheme in that
a class of
potentials is found to exist which share the same real energy eigenvalues
and have
their eigenfunctions derived from an application of the sl(2,\C) generators on
normalized states.\par
%
%
\section{sl(2, {\sixteenof C}) algebra and its realization}

Let us begin by noting that the commutation relations of sl(2,\C) (or $A_1$ in
Cartan's classification of simple complex Lie algebras), namely
\begin{equation}
  [J_0, J_{\pm}] = \pm J_{\pm}, \qquad [J_+, J_-] = - 2J_0,  \label{eq:alg}
\end{equation}
can be given a differential realization
\begin{equation}
  J_0 = - {\rm i} \frac{\partial}{\partial\phi}, \qquad J_{\pm} =
e^{\pm{\rm i}\phi}
  \left[\pm\frac{\partial}{\partial x} + \left({\rm i}
\frac{\partial}{\partial\phi}
  \mp \frac{1}{2}\right) F(x) + G(x)\right],  \label{eq:J}
\end{equation}
where the auxiliary variable $\phi$ ranges in $0 \le \phi < 2\pi$, $x \in
\R$, and
the two functions $F(x)$, $G(x) \in \C$ satisfy coupled differential equations
\begin{equation}
  \frac{dF}{dx} = 1 - F^2, \qquad \frac{dG}{dx} = - FG.  \label{eq:diff-eq}
\end{equation}
Note that since $J_- \ne J_+^{\dagger}$, we generate an sl(2,\C) algebra rather
than so(2,1), which is consistent with $J_- = J_+^{\dagger}$.\par
%
%
The Casimir operator corresponding to the above generators is
\begin{equation}
  J^2 = J_0^2 \mp J_0 - J_{\pm} J_{\mp}.
\end{equation}
In terms of $F$ and $G$, it reads
\begin{equation}
  J^2 = \frac{\partial^2}{\partial x^2} - \left(\frac{\partial^2}{\partial
\phi^2} +
  \frac{1}{4}\right) F' + 2 {\rm i} \frac{\partial}{\partial \phi} G' - G^2
- \frac{1}{4},
  \label{eq:casimir}
\end{equation}
where a prime denotes derivative with respect to spatial variable $x$.\par
%
%
In the so(2,1) case~\cite{alhassid, wu, englefield}, one considers for
bound states
unitary irreducible representations of type $D^+_k$, for which
\begin{eqnarray}
  J_0 |km\rangle & = & m |km\rangle, \qquad m=k, k+1, k+2, \ldots, \nonumber \\
  J^2 |km\rangle & = & k(k-1) |km\rangle,  \label{eq:rep}
\end{eqnarray}
and $k$ is positive (but not necessarily restricted to integers or half-integers
as one only deals with the algebra and not with the whole group~\cite{barut}).
When going to the complex Lie algebra sl(2,\C), Eq.~(\ref{eq:rep}) still defines
irreducible representations. These are the only ones we shall consider here.\par
%
%
Hence, in the remainder of this paper, we are going to look for solutions
of~(\ref{eq:rep}) given by
\begin{equation}
  |km\rangle = \Psi_{km}(x, \phi) = \psi_{km}(x) \frac{e^{{\rm i}
  m\phi}}{\sqrt{2\pi}}, \label{eq:Psi}
\end{equation}
where $k > 0$ and $m = k+n$, $n=0$, 1, 2,~\ldots. From
Eqs.~(\ref{eq:casimir})--(\ref{eq:Psi}), it follows that the coefficient
functions
$\psi_{km}(x)$ obey the Schr\"odinger equation
\begin{equation}
  - \psi''_{km} + V_m \psi_{km} = - \left(k - \case{1}{2}\right)^2 \psi_{km}.
  \label{eq:SE}
\end{equation}
In (\ref{eq:SE}), the one-parameter family of potentials, denoted $V_m$, is
represented by
\begin{equation}
  V_m = \left(\case{1}{4} - m^2\right) F' + 2m G' + G^2.  \label{eq:Vm}
\end{equation}
Provided they are normalizable, the functions $\psi_{km}(x)$ are the $(n+1)$th
bound state wavefunctions for the potentials $V_m$, corresponding to the energy
eigenvalues
\begin{equation}
  E^{(m)}_n = - \left(m - n - \case{1}{2}\right)^2.  \label{eq:Em}
\end{equation}
The potentials $V_m$, $m=k$, $k+1$, $k+2$,~\ldots, supporting the same
eigenvalues $- \left(k - \frac{1}{2}\right)^2$ produce a potential algebra:
$J_+$ and $J_-$ connect among themselves all the wavefunctions corresponding to
the same energy, but to different potentials $V_m$.\par
%
%
Before we embark upon our detailed study of sl(2,\C), let us make a few
remarks on
the use of~(\ref{eq:alg}) as an so(2,1) potential algebra in the real
domain. Wu and
Alhassid~\cite{wu} showed that Morse, P\"oschl-Teller, and Rosen-Morse
potentials emerge as particular solutions of~(\ref{eq:diff-eq}). Later
Englefield and
Quesne~\cite{englefield} quite exhaustively identified three classes of so(2,1)
solutions for the coupled set of equations~(\ref{eq:diff-eq}) according as
$F^2 < 1$,
$F^2 > 1$, or $F^2 = 1$:
\begin{equation}
\begin{array}{lll}
      F^2<1: & F(x) = \tanh(x-c), & G(x) = b \sech(x-c), \\[0.2cm]
      F^2>1: & F(x) = \coth(x-c), & G(x) = b \cosech(x-c), \\[0.2cm]
      F^2=1: & F(x) = \pm 1, & G(x) = b e^{\mp x},
\end{array}  \label{eq:3cases}
\end{equation}
where $b$ and $c$ are real constants. When substituted in~(\ref{eq:Vm}),
the above
possibilities for $F$ and $G$ lead to three distinct types of potentials.
These are
the (nonsingular) Scarf II or first Gendenshtein potential~\cite{gendenshtein}
$V_1$, given by
\begin{equation}
  V_1(x) = \left(b^2 - m^2 + \case{1}{4}\right) \sech^2 x - 2mb \sech x \tanh x,
  \label{eq:V1}
\end{equation}
the (singular) generalized P\"oschl-Teller or second Gendenshtein
potential~\cite{gendenshtein} $V_2$, given by
\begin{equation}
  V_2(x) = \left(b^2 + m^2 - \case{1}{4}\right) \cosech^2 x - 2mb \cosech x
\coth x,
  \label{eq:V2}
\end{equation}
and the Morse potential $V_3$, given by
\begin{equation}
  V_3(x) = b^2 e^{-2x} - 2mb e^{-x}.  \label{eq:V3}
\end{equation}
Eqs.~(\ref{eq:V1}) and~(\ref{eq:V2}) correspond to $c=0$, while
Eq.~(\ref{eq:V3})
corresponds to the upper signs in~(\ref{eq:3cases}). These simplifications
do not
alter the physical significance of the results. Note that our convention of
denoting
the potentials differs slightly from Ref.~\cite{englefield}, where the Morse
potential is referred to as $V_2$ and the generalized P\"oschl-Teller or second
Gendenshtein potential as $V_3$.\par
%
%
\section{General results for complex potentials associated with sl(2,
{\sixteenof
C})}
We now proceed to write down solutions of Eq.~(\ref{eq:diff-eq}). This can
be done
either by splitting the functions $F$ and $G$ into their real and imaginary
components and looking for solutions of the four real equations satisfied
by them
or, equivalently, solving for the first equation of~(\ref{eq:diff-eq})
directly to find
$F$ and then substituting it in the second equation of~(\ref{eq:diff-eq}) to
determine $G$. For simplicity we choose the second approach. Our solutions are
summarized as follows:
\begin{equation}
\begin{array}{lll}
      {\rm I}: & F(x) = \tanh(x-c-{\rm i}\gamma), & G(x) = (b_R+{\rm i}b_I)
            \sech(x-c-{\rm i}\gamma), \\[0.2cm]
      {\rm II}: & F(x) = \coth(x-c-{\rm i}\gamma), & G(x) = (b_R+{\rm i}b_I)
            \cosech(x-c-{\rm i}\gamma), \\[0.2cm]
      {\rm III}: & F(x) = \pm 1, & G(x) = (b_R+{\rm i}b_I) e^{\mp x},
\end{array}
\end{equation}
where $b = b_R + {\rm i} b_I$, $b_R$, $b_I \in \R$, and $-\frac{\pi}{4} \le
\gamma <
\frac{\pi}{4}$.\par
%
%
The resulting potentials are given by
\begin{eqnarray}
  {\rm I:\quad} V_m & = & \left[(b_R + {\rm i}b_I)^2 - m^2 + \case{1}{4}\right]
         \sech^2(x - c - {\rm i}\gamma) \nonumber \\
  && \mbox{} - 2m (b_R + {\rm i}b_I) \sech(x - c - {\rm i}\gamma)
         \tanh(x - c - {\rm i}\gamma),  \label{eq:VmI} \\
  {\rm II:\quad} V_m & = & \left[(b_R + {\rm i}b_I)^2 + m^2 - \case{1}{4}\right]
         \cosech^2(x - c - {\rm i}\gamma) \nonumber \\
  && \mbox{} - 2m (b_R + {\rm i}b_I) \cosech(x - c - {\rm i}\gamma)
         \coth(x - c - {\rm i}\gamma),  \label{eq:VmII} \\
  {\rm III:\quad} V_m & = & (b_R + {\rm i}b_I)^2 e^{\mp 2x} \mp 2m (b_R + {\rm
         i}b_I) e^{\mp x}.  \label{eq:VmIII}
\end{eqnarray}
The potentials (\ref{eq:VmI})--(\ref{eq:VmIII}) can be looked upon as the
complexified versions of the corresponding real ones
(\ref{eq:V1})--(\ref{eq:V3}).\par
%
%
By separating out the real and imaginary parts, these complex potentials can be
expressed as
\begin{eqnarray}
  {\rm I:\quad} V_m & = & \frac{2}{[\cosh 2(x-c) + \cos 2\gamma]^2} \Bigl\{
\Bigl(
          b_R^2 - b_I^2 - m^2 + \case{1}{4}\Bigr) \Bigl[1 + \cosh 2(x-c)
\cos 2\gamma
          \Bigr]\nonumber \\
  && \mbox{} - 2 b_R b_I \sinh 2(x-c) \sin 2\gamma \nonumber \\
  && \mbox{} - 2m \Bigl[b_R \sinh(x-c) \cos\gamma \Bigl(\cosh 2(x-c) - \cos
          2\gamma + 2\Bigr) \nonumber \\
  && \mbox{} - b_I \cosh(x-c) \sin \gamma \Bigl(\cosh 2(x-c) - \cos 2\gamma - 2
          \Bigr)\Bigr]\Bigr\} \nonumber \\
  && \mbox{} + \frac{2{\rm i}}{[\cosh 2(x-c) + \cos 2\gamma]^2} \Bigl\{ \Bigl(
          b_R^2 - b_I^2 - m^2 + \case{1}{4}\Bigr) \sinh 2(x-c) \sin 2\gamma
          \nonumber \\
  && \mbox{} + 2 b_R b_I \Bigl[1 + \cosh 2(x-c) \cos 2\gamma\Bigr] \nonumber \\
  && \mbox{} - 2m \Bigl[b_R \cosh(x-c) \sin\gamma \Bigl(\cosh 2(x-c) - \cos
          2\gamma - 2\Bigr)\nonumber \\
  && \mbox{} + b_I \sinh(x-c) \cos \gamma \Bigl(\cosh 2(x-c) - \cos 2\gamma + 2
          \Bigr)\Bigr]\Bigr\} \nonumber \\
  {\rm II:\quad} V_m & = & \frac{2}{[\cosh 2(x-c) - \cos 2\gamma]^2}
\Bigl\{ \Bigl(
          b_R^2 - b_I^2 + m^2 - \case{1}{4}\Bigr) \Bigl[- 1 + \cosh 2(x-c) \cos
          2\gamma\Bigr] \nonumber \\
  && \mbox{} - 2 b_R b_I \sinh 2(x-c) \sin 2\gamma \nonumber \\
  && \mbox{} - 2m \Bigl[b_R \cosh(x-c) \cos\gamma \Bigl(\cosh 2(x-c) + \cos
          2\gamma - 2\Bigr)\nonumber \\
  && \mbox{} - b_I \sinh(x-c) \sin \gamma \Bigl(\cosh 2(x-c) + \cos 2\gamma + 2
          \Bigr)\Bigr]\Bigr\} \nonumber \\
  && \mbox{} + \frac{2{\rm i}}{[\cosh 2(x-c) - \cos 2\gamma]^2} \Bigl\{ \Bigl(
          b_R^2 - b_I^2 + m^2 - \case{1}{4}\Bigr) \sinh 2(x-c) \sin 2\gamma
          \nonumber \\
  && \mbox{} + 2 b_R b_I \Bigl[- 1 + \cosh 2(x-c) \cos 2\gamma\Bigr]\nonumber \\
  && \mbox{}  - 2m \Bigl[b_R \sinh(x-c) \sin \gamma \Bigl(\cosh 2(x-c) + \cos
          2\gamma + 2\Bigr)\nonumber \\
  && \mbox{} + b_I \cosh(x-c) \cos \gamma \Bigl(\cosh 2(x-c) + \cos 2\gamma - 2
          \Bigr)\Bigr]\Bigr\} \nonumber \\
  {\rm III:\quad} V_m & = & \left(b_R^2 - b_I^2\right) e^{\mp 2x} \mp 2m b_R
          e^{\mp x} + {\rm i} b_I \left(2 b_R e^{\mp 2x} \mp 2m e^{\mp
x}\right).
          \label{eq:comp-Vm}
\end{eqnarray}
\par
%
%
The above categories of potentials are displayed in their most general forms and
give a quite complete realization of sl(2,\C) algebra corresponding to the
representation~(\ref{eq:J}). It should be remarked that although $V_m$ of
(II) can
be apparently obtained from $V_m$ of (I) by the transformation $\gamma \to
\gamma \pm \frac{\pi}{2}$, $b_R \to \pm b_I$, $b_I \to \mp b_R$, the range
of the
parameter $\gamma$ is not left invariant. So the potentials (I) and (II)
are indeed
different and quite independent  of one another. Note further that the
special case
$\gamma = b_I = 0$ reduces the three potentials of~(\ref{eq:comp-Vm}) to
their real forms (\ref{eq:V1})--(\ref{eq:V3}), as it should be. On the
other hand, the
case $\gamma = b_R = 0$ (that is $F \in \R$, $G \in {\rm i} \R$) gives
\begin{eqnarray}
  {\rm I:\quad} V_m & = & \left(- b_I^2 - m^2 + \case{1}{4}\right)
\sech^2(x-c) -
          2 {\rm i} m b_I \sech(x-c) \tanh(x-c),  \label{eq:VmI-spe} \\
  {\rm II:\quad} V_m & = & \left(- b_I^2 + m^2 - \case{1}{4}\right)
\cosech^2(x-c) -
          2 {\rm i} m b_I \cosech(x-c) \coth(x-c), \\
  {\rm III:\quad} V_m & = & - b_I^2 e^{\mp 2x} \mp 2 {\rm i} m b_I e^{\mp x}.
\end{eqnarray}
Obviously PT symmetry holds for I, but not for the other two.\par
%
%
\section{Analysis of some special cases}
\subsection{Complexification of the Scarf II potential}
In the literature~\cite{levai, cooper}, the real Scarf II potential is
usually given in
the form
\begin{equation}
  V^{(S)}(x) = \left[B^2 - A (A+1)\right] \sech^2 x + B (2A+1) \sech x \tanh x.
  \label{eq:VS}
\end{equation}
This potential is well known to be exactly solvable. For $A>0$, the associated
eigenfunctions and eigenvalues are
\begin{eqnarray}
  \psi_n(x) & = & N_n (\sech x)^A \exp[- B \arctan(\sinh x)] P_n^{(- {\rm
i} B - A -
          \frac{1}{2}, {\rm i} B - A - \frac{1}{2})} ({\rm i} \sinh x),
\label{eq:wfS} \\
  E_n & = & - (A-n)^2, \qquad n=0, 1, \ldots, n_{\rm max} < A,  \label{eq:ES}
\end{eqnarray}
where $N_n$ is a normalization constant, and $P_n^{(\alpha,\beta)}$ is a Jacobi
polynomial.\par
%
%
Replacing $B$ by ${\rm i} B$ leads to the PT-symmetric form of the
potential~(\ref{eq:VS}):
\begin{equation}
  V^{(CS)}(x) = - \left[B^2 + A (A+1)\right] \sech^2 x + {\rm i} B (2A+1)
\sech x
  \tanh x,  \label{eq:VCS}
\end{equation}
where the real and imaginary parts have no singularity on the real axis.
Further,
they are invariant under the exchange $A + \frac{1}{2} \leftrightarrow B$.
Without
loss of generality, we may  assume $A + \frac{1}{2} > 0$ along with $B >
0$, since
replacing $B$ by $-B$ only changes $V$ into $V^*$ with both $V$ and $V^*$
bearing
the same real $E_n$ corresponding to the wavefunctions $\psi_n(x)$ and
$\psi_n^*(x)$, respectively.\par
%
%
Comparing (\ref{eq:VCS}) with (\ref{eq:VmI-spe}) obtained from the sl(2,\C)
algebra, we find $c=0$ and
\begin{eqnarray}
  b_I^2 + m^2 - \case{1}{4} & = & B^2 + A (A+1), \label{eq:cond1} \\
  - 2m b_I & = & B (2A+1). \label{eq:cond2}
\end{eqnarray}
While (\ref{eq:cond2}) gives $m = - B (2A+1)/(2b_I)$, using it
in~(\ref{eq:cond1})
gives two solutions for $b_I^2$ as
\begin{equation}
  b_I^2 = B^2, \qquad b_I^2 = \left(A + \case{1}{2}\right)^2.
\end{equation}
If $b_I^2 = B^2$, then $b_I = \epsilon B$ and $m = - \epsilon \left(A +
\frac{1}{2}\right)$, where $\epsilon = \pm 1$. Since by assumption $m > 0$ (see
discussion in Section~2), we have to choose $\epsilon = -1$ (to be
consistent with
$A + \frac{1}{2} > 0$) implying $b_I = - B$, $m = A + \frac{1}{2}$. On the
other hand,
if $b_I^2 = \left(A + \case{1}{2}\right)^2$, then $b_I = \epsilon \left(A +
\frac{1}{2}\right)$ and $m = - \epsilon B$, where we have again to choose
$\epsilon
= -1$ (to be consistent with $B>0$), so that $b_I = - \left(A +
\frac{1}{2}\right)$,
$m = B$. We thus get  two (noncommuting) sl(2,\C) algebras, which can be mapped
onto each other by effecting a transformation $A + \frac{1}{2}
\leftrightarrow B$.
Let us denote their generators by $J_0^{(i)}$, $J_+^{(i)}$, $J_-^{(i)}$,
$i=1$, 2,
where $i=1$ (resp.~2) corresponds to $(m, b_I) = \left(A + \frac{1}{2}, -
B\right)$
[resp.\ $\left(B, - A - \frac{1}{2}\right)$].\par
%
%
The eigenfunctions corresponding to~(\ref{eq:VCS}) can be arrived at by using in
turn both algebras. For the algebra labelled by $i$, the wavefunction
corresponding
to $n=0$ follows from $J_-^{(i)} \Psi_{kk}^{(i)} = 0$, while those for $n
\ne 0$ are
obtained from the latter by employing the property $J_+^{(i)} \Psi_{km}^{(i)} =
\alpha_{km} \Psi_{k,m+1}^{(i)}$, where $\alpha_{km}$ are some constants. For
conciseness, we do not give the details of our calculations, which will be
presented
elsewhere. We just state our results, which are
\begin{equation}
  \psi_n^{(m)}(x) = N_n^{(m)} (\sech x)^{m - \frac{1}{2}} \exp[{\rm i} b_I
  \arctan(\sinh x)] P_n^{(- b_I - m, b_I - m)} ({\rm i} \sinh x),
\label{eq:wfCS}
\end{equation}
where $(m, b_I) = \left(A + \frac{1}{2}, - B\right)$ or $\left(B, - A -
\frac{1}{2}\right)$. The corresponding eigenvalues are given
by~(\ref{eq:Em}).\par
%
%
The functions~(\ref{eq:wfCS}) being normalizable on the real line, we
conclude that
there is in general a doubling of energy levels when going from the real to the
complex Scarf II potential. The latter indeed has two series of levels
associated
with the two algebras: the first ones,
\begin{equation}
  E_n^{(A + \frac{1}{2})} = - (A - n)^2, \qquad n=0, 1, \ldots, n^{(A +
  \frac{1}{2})}_{\rm max} < A,  \label{eq:ECS1}
\end{equation}
coincide with the levels~(\ref{eq:ES}) of the real potential, with the
corresponding
wavefunctions obtained from~(\ref{eq:wfS}) by the substitution $B \to {\rm
i} B$,
while the second ones,
\begin{equation}
  E_{n}^{(B)} = - \left(B - n - \case{1}{2}\right)^2, \qquad n=0, 1, \ldots,
  n^{(B)}_{\rm max} < B - \case{1}{2},  \label{eq:ECS2}
\end{equation}
have no counterparts in the real case. Note that only the first set of
wavefunctions
was mentioned in the brief account of the complexified Scarf II potential
made in
Ref.~\cite{znojil00}.\par
%
%
Whenever $|A + \frac{1}{2} - B|$ approaches an integer $m$, some levels
corresponding to~(\ref{eq:ECS1}) and~(\ref{eq:ECS2}) become quasi-degenerate. It
can be checked that for $|A + \frac{1}{2} - B| = m$, the corresponding
wavefunctions~(\ref{eq:wfCS}) become proportional, so that we again observe the
phenomenon of unavoided level crossings without degeneracy, previously
encountered for the PT-symmetric harmonic oscillator~\cite{znojil99a}.\par
%
%
It is of interest to consider the special case $B=1$. We obtain
from~(\ref{eq:VCS})
\begin{equation}
  V = - \left[\left(A + \case{1}{2}\right)^2 + \case{3}{4}\right] \sech^2 x
+ 2 {\rm i}
  \left(A + \case{1}{2}\right) \sech x \tanh x.  \label{eq:VCSspe}
\end{equation}
It can be immediately seen that by setting $A + \frac{1}{2} = - \lambda$,
($\lambda
< 0$), (\ref{eq:VCSspe}) reduces to the potential $V^{(1)} - \frac{1}{4}$ of
Ref.~\cite{bagchi00a} for $\mu=1$. The energies~(\ref{eq:ECS1}) obtained
from the
first sl(2,\C) algebra become $E_n^{(-\lambda)} = - \left(\lambda + n +
\frac{1}{2}\right)^2$ and coincide with $E_n^{(2)} - \frac{1}{4}$ of Eq.~(6)
in~\cite{bagchi00a}. The second algebra leads to a single energy level
corresponding to~(\ref{eq:ECS2}) for $B=1$ and $n=0$, that is $E_0^{(1)} = -
\frac{1}{4}$, which is consistent with the zero-energy state
of~\cite{bagchi00a}.\par
%
%
\subsection{Complexification of the generalized P\"oschl-Teller potential}
The wavefunctions and energy levels of the complexified generalized
P\"oschl-Teller potential~(\ref{eq:VmII}) with $b_I=0$ can be found out in a
manner similar to the one of~(\ref{eq:VmI}) with $\gamma = b_R = 0$. Again we
reserve the details of our calculations for a future communication. Let us
however
mention two interesting aspects of the potential.\par
%
%
Writing it in a form similar to its real counterpart~\cite{levai, cooper},
we obtain
\begin{equation}
  V^{(CGPT)}(x) = \left[B^2 + A (A+1)\right] \cosech^2(x - {\rm i} \gamma) - B
  (2A+1) \cosech(x - {\rm i} \gamma) \coth(x - {\rm i} \gamma).
\label{eq:VCGPT}
\end{equation}
The real potential, corresponding to $\gamma=0$, being singular must be confined
to the semi-axis $(0, +\infty)$. The complexified potential~(\ref{eq:VCGPT}) gets
regularized by the complex shift $x \to x - {\rm i} \gamma$ and may be
considered
on the whole real line. Using now a complex analogue of the point canonical
coordinate transformation known to relate the generalized P\"oschl-Teller and
P\"oschl-Teller II potentials~\cite{levai,cooper}, the
potential~(\ref{eq:VCGPT})
can be changed into the complexified P\"oschl-Teller II potential
\begin{equation}
  V^{(CPT)}(t) = \frac{(B-A) (B-A-1)}{\sinh^2(t - {\rm i}\epsilon)} -
\frac{(A+B)
  (A+B+1)}{\cosh^2(t - {\rm i}\epsilon)},  \label{eq:VCPT}
\end{equation}
where $t = x/2$ and $\epsilon = \gamma/2$. In Ref.~\cite{znojil99b}, the real
spectrum and corresponding wavefunctions of the potential~(\ref{eq:VCPT}) were
found by solving the Schr\"odinger equation. It is remarkable that they can
also be
derived algebraically by applying the above-mentioned complex point canonical
coordinate transformation to the sl(2,\C) results for the
potential~(\ref{eq:VCGPT}).\par
%
%
On the other hand, some time ago Andrianov {\sl et al.}~\cite{andrianov} wrote
down a transparent complex potential in the form
\begin{equation}
  V^{(1)}(y) = \frac{2\epsilon_R}{\cosh^2[\sqrt{-\epsilon_R} (y+b) + {\rm
i} \rho]},
  \label{eq:Vtransp}
\end{equation}
where $\epsilon_R < 0$, $b \in \R$, $\rho = \pm \frac{1}{2} \arctan \left(2
\sqrt{-\epsilon_R}/a\right) \ne \frac{\pi}{2} (2n+1)$, and $a \in \R_0$.
$V^{(1)}(y)$
is invariant under PT with its imaginary part being P odd. It vanishes at
infinity
and has a bound state with energy~$\epsilon_R$. Furthermore it can be obtained
from the real transparent potential $\sech^2x$ by a complex shift of the
coordinate~$x$.\par
%
%
The potential~(\ref{eq:Vtransp}) compares well with~(\ref{eq:VCGPT}) with $B =
A+1$ (and $x \to x-c$):
\begin{equation}
  V^{(CGPT)}(x) = - \case{1}{2} (A+1) (2A+1) \sech^2 \case{1}{2} (x-c-{\rm
  i}\gamma).  \label{eq:VCGPTspe}
\end{equation}
Indeed under a change of variable $y = x/(2\sqrt{-\epsilon_R})$, the Hamiltonian
corresponding to~(\ref{eq:Vtransp}) reads
\begin{equation}
  H = - 4\epsilon_R \left[- \frac{d^2}{dx^2} - \case{1}{2} \sech^2 \case{1}{2}
  (x-c-{\rm i}\gamma)\right],  \label{eq:Htransp}
\end{equation}
where we have put $c = - 2 \sqrt{-\epsilon_R}\, b$ and $\gamma = - 2\rho$. Apart
from a multiplicative constant, the above Hamiltonian is a particular case of
\begin{equation}
  H^{(CGPT)} = - \frac{d^2}{dx^2} - \case{1}{2} (A+1) (2A+1) \sech^2 \case{1}{2}
  (x-c-{\rm i}\gamma)  \label{eq:HCGPTspe}
\end{equation}
[obtained from~(\ref{eq:VCGPTspe})] for $A=0$. From the algebraic results
for the
general potential~(\ref{eq:VCGPT}), it follows that the
Hamiltonian~(\ref{eq:HCGPTspe}) with $A=0$ has a single bound state with energy
$-\frac{1}{4}$, thus giving a single bound state with energy $\epsilon_R$
for the
Hamiltonian~(\ref{eq:Htransp}), as it should be.\par
%
%
\subsection{Complexification of the Morse potential}
As a last example, we consider the complexified Morse potential
\begin{equation}
  V^{(CM)}(x) = (B_R + {\rm i} B_I)^2 e^{-2x} - (B_R + {\rm i} B_I) (2A+1)
e^{-x},
  \label{eq:VCM}
\end{equation}
corresponding to the potential~(\ref{eq:VmIII}) and giving back the real
one~\cite{levai, cooper} for $B_I=0$. It is straightforward to show that
for $A$ and
$B_R$ positive, the potential~(\ref{eq:VCM}) has the same real eigenvalues
as its
real counterpart (but of course different wavefunctions). This provides a very
simple example of non-PT-symmetric complex potential with real eigenvalues.
Other examples of such potentials are already known (see
e.g.~\cite{cannata}).\par
%
%
\section{Conclusion}
In this letter, we showed that the powerful group theoretical formalism of
potential algebras can be extended to non-Hermitian Hamiltonians by a simple
complexification of the real algebras considered for Hermitian
Hamiltonians. This
provides us with a simple method of constructing both PT-symmetric and
non-PT-symmetric Hamiltonians with real eigenvalues. We considered here the case
of the sl(2,\C) (or $A_1$) potential algebra, corresponding to the
complexification
of so(2,1).\par
%
%
Our construction method of new non-Hermitian Hamiltonians with real eigenvalues
may be extended in two ways. First, instead of complexifying so(2,1), we
may try to
complexify the larger algebra $\mbox{so(2,2)} \simeq \mbox{so(2,1)} \oplus
\mbox{so(2,1)}$, which is a potential algebra for the whole class of real
Natanzon
potentials~\cite{alhassid, wu}. Second, to all the potentials constructed
by using
either sl(2,\C) or some generalization thereof, we may apply complex
analogues of
the transformations used to inter-relate real exactly solvable
potentials~\cite{levai, cooper}, as we did in Section~4.2. It should be
clear that
both of these extensions will give rise to a whole menagerie of new Hamiltonians
with real eigenvalues.\par
%
%


\newpage
\begin{thebibliography}{99}

\bibitem{ballentine} L.M.\ Ballentine, Quantum Mechanics - A Modern Development
(World Scientific, Singapore, 1998).

\bibitem{alhassid} Y.\ Alhassid, F.\ G\"ursey, F.\ Iachello, Ann.\ Phys.\
(N.Y.) 148
(1983) 346; 167 (1986) 181;\\
J.\ Wu, Y.\ Alhassid, F.\ G\"ursey, Ann.\ Phys.\ (N.Y.) 196 (1989) 163.

\bibitem{bohm} A.\ Bohm, Y.\ Ne'eman, A.O.\ Barut, Dynamical Groups and Spectrum
Generating Algebras (World Scientific, Singapore, 1988).

\bibitem{bender98a} C.M.\ Bender, S.\ Boettcher, Phys.\ Rev.\ Lett.\ 80 (1998)
5243.

\bibitem{bender98b} C.M.\ Bender, S.\ Boettcher, J.\ Phys.\ A 31 (1998) L273.

\bibitem{cannata} F.\ Cannata, G.\ Junker, J.\ Trost, Phys.\ Lett.\ A 246 (1998)
219.

\bibitem{andrianov} A.A.\ Andrianov, M.V.\ Ioffe, F.\ Cannata, J.-P.\
Dedonder, Int.\
J.\ Mod.\ Phys.\ A 14 (1999) 2675.

\bibitem{znojil99a} M.\ Znojil, Phys.\ Lett.\ A 259 (1999) 220.

\bibitem{bagchi00a} B.\ Bagchi, R.\ Roychoudhury, J.\ Phys.\ A 33 (2000) L1.

\bibitem{znojil00} M.\ Znojil, J.\ Phys.\ A 33 (2000) L61.

\bibitem{bagchi00b} B.\ Bagchi, F.\ Cannata, C.\ Quesne, Phys.\ Lett.\ A 269
(2000) 79.

\bibitem{znojil99b} M.\ Znojil, New set of exactly solvable complex potentials
giving the real energies, preprint quant-ph/9912079.

\bibitem{feshbach} H.\ Feshbach, C.E.\ Porter, V.F.\ Weisskopf, Phys.\ Rev.\ 96
(1954) 448.

\bibitem{wu} J.\ Wu, Y.\ Alhassid, J.\ Math.\ Phys.\ 31 (1990) 557.

\bibitem{englefield} M.J.\ Englefield, C.\ Quesne, J.\ Phys.\ A 24 (1991) 3557.

\bibitem{barut} A.O.\ Barut, C.\ Fronsdal, Proc.\ Roy.\ Soc.\ (London) A
287 (1965)
532.

\bibitem{gendenshtein} L.E.\ Gendenshtein, JETP Lett.\ 38 (1983) 356.

\bibitem{levai} G.\ L\'evai, J.\ Phys.\ A 22 (1989) 689; 27 (1994) 3809.

\bibitem{cooper} F.\ Cooper, A.\ Khare, U.\ Sukhatme, Phys.\ Rep.\ 251
(1995) 267.

\end {thebibliography}
\end{document}